\documentclass[preprint]{elsarticle}
\usepackage{lineno,hyperref,pdflscape}
\modulolinenumbers[5]

\usepackage{multicol}
\usepackage{color}
\usepackage{xspace}
\usepackage{enumerate}
\usepackage{amsmath} 
\usepackage{ulem} 
\usepackage{placeins}
\usepackage{threeparttable}

\usepackage{tabularx}
\usepackage{mathtools} 
\usepackage{amsfonts}
\newcolumntype{M}{>{\centering\arraybackslash}m{1.85cm}}
\usepackage[export]{adjustbox}
\usepackage{float}
\usepackage{braket}
\usepackage{lipsum}
\usepackage{longtable}
\graphicspath{{figure/}}
\newcommand\T{\rule{0pt}{3ex}}       
\newcommand\B{\rule[-1.5ex]{0pt}{0pt}} 

\usepackage{booktabs}
\usepackage{blindtext}
\makeatletter
\newcommand{\colorcaption}[2][]{%
	\begingroup%
	\renewcommand{\@caption@fignum@sep}{ (Color online). }%
	\caption[#1]{#2}%
	\endgroup%
}
\makeatother

\usepackage{amssymb}
\usepackage{graphicx}
\usepackage{rotating}
\usepackage{tikz}
\begin{document}
	
\begin{frontmatter}
\title{\textit{Ab initio} no-core shell model study of neutron-rich $^{18,19,20}$C isotopes}
\author{Priyanka Choudhary\footnote{pchoudhary@ph.iitr.ac.in}}
\author{Praveen C. Srivastava\footnote{Corresponding author: praveen.srivastava@ph.iitr.ac.in}}
\address{Department of Physics, Indian Institute of Technology Roorkee, Roorkee 247667, India}
\date{\hfill \today}
\begin{abstract}
We implement the \textit{ab initio} no-core shell model approach to study neutron rich  $^{18}$C, $^{19}$C and $^{20}$C isotopes. For this purpose, we employ charge-dependent Bonn 2000 (CDB2K), inside non-local outside Yukawa (INOY) and chiral next-to-next-to-next-to-leading order (N$^{3}$LO) nucleon-nucleon interactions. Low-lying energy spectra, electromagnetic properties and point-proton radii for these nuclei up to basis space $N_{\rm max}$ = 4 are calculated. Binding energies obtained with INOY interaction are in better agreement with the experimental values as compared to other \textit{ab initio} interactions. We also show the behavior of ground state energy and point-proton radii with the NCSM parameters, $\hbar$$\Omega$ and $N_{\rm max}$.
We report a strong sensitivity of the B(E2) values from the first excited $2^+$ to ground state of $^{18}$C and $^{20}$C to the nuclear interaction. Shell model calculations with YSOX interaction are also performed, and corresponding results are compared with 
\textit{ab initio} one. 
\end{abstract}
		
		
\end{frontmatter}
	
	

\section{Introduction}
The study of the nuclear structure of atomic nuclei far from the beta stability line poses a challenge to theory as well as experiments due to their exotic behavior, such as halo structure \cite{I.Tanihata,Hansen} and change in the shell structure \cite{O.Sarlin,T.Otsuka1,T.Otsuka2}. The neutron halo nuclei are characterized by weakly bound neutron(s), which are spatially decoupled from the nuclear core leading to extended core-decoupled wave functions (extended radial distribution). Ongoing development in the experimental facilities promotes the study of nuclei near to drip line.

Neutron-rich carbon isotopes are of interest due to their exotic features across the isotopic chain. For instance, $^{19}$C is a one-neutron halo nucleus \cite{19C}, and $^{22}$C, which is a drip line nuclei, is also a candidate for a two-neutron halo nucleus \cite{22C}. The radioactive nuclear beam facilities at GANIL, MSU, GSI, and RIKEN have been used to study the structural properties of near drip line carbon isotopes. The reduced electric quadrupole transition strength [B(E2)] is an essential fundamental observable which provides information about the weakly bound and decoupled neutrons from the core. The B(E2) from the first $2^+$ state to the ground state (g.s.) has been measured through several experiments over the past years for even-even carbon isotopes  and different values are reported for the same nucleus, obtained from different measurements \cite{ADNDT}. Imai \textit{et al.} \cite{Imai} have measured extremely quenched B(E2) value 0.63 $\rm e^2 \rm fm^4$ for $^{16}$C and claimed that this is due to increase of shell gap between proton $p_{3/2}$ and $p_{1/2}$ orbitals from $^{14}$C ($N=8$) to $^{16}$C ($N=10$). However, based on the measurement by Wiedeking \textit{et al} \cite{Wiedeking}, larger B(E2) value 4.15(73) $\rm e^2 \rm fm^4$ is derived for $^{16}$C and after reanalyzing the data of Ref. \cite{Imai}, a larger but still quenched B(E2)=  2.6 $\pm$ 0.9 $\rm e^2 \rm fm^4$ was obtained \cite{Ong}. Both values of B(E2) do not favor the decoupling of valence neutron from the core. The B(E2) ($4.21 \substack{+0.34\\-0.26}(\rm stat) \substack{+0.28\\-0.24}(\rm syst_{B\rho}) \substack{+0.64\\-0.00}(\rm syst_{\rm feeding})$ $\rm e^2 \rm fm^4$), obtained from a new lifetime measurement \cite{Petri} agrees well with the previous measurement \cite{Wiedeking}. No-core shell model (NCSM) \cite{Phys.Rev.C542986,Phys.Rev.C573119,PRL84,2000,PRC62,2009,MVS2009,nocore_review,stetcu1,stetcu2,Forssen} is an \textit{ab initio} microscopic approach which has successfully described properties of lighter nuclei. The NCSM calculations with several \textit{NN} and \textit{NN}+\textit{NNN} interactions have been carried out for $^{16}$C in Ref. \cite{Forssen}.  The extrapolated B(E2) value is underestimated for $^{16}$C with the used CDB2K interaction. Energy levels of $^{16}$C are also calculated, and the values obtained with the CDB2K and chiral N$^3$LO \textit{NN} are in reasonable agreement with the experiment. Inclusion of \textit{NNN} interaction further improves the excitation energies of the states. Similarly, different claims have been made based on the different experimental results for $^{18}$C and $^{20}$C. Theoretical calculations with shell model \cite{SHF,20C_2011,neutronrichC}, antisymmetrized molecular dynamics (AMD) \cite{AMD}, the multi-Slater determinant AMD (AMD+MSD) \cite{AMD+MSD}, deformed Skyrme Hartree-Fock \cite{SHF} and NCSM \cite{Forssen} predicted different values of B(E2) for these nuclei  in the past. Motivated by these experimental data and different theoretical claims, we have performed \textit{ab initio} NCSM calculations for $^{18-20}$C isotopes using three realistic interactions \textit{i.e.} CDB2K, INOY, and N\textsuperscript{3}LO. Apart from the \textit{ab initio} calculations, we have carried out shell model calculations in $psd$-model space for comparison. Earlier, Forss{\'e}n \textit{et al.} \cite{Forssen} have studied even-even carbon isotopes in the range $A = 10-20$ within the framework of the NCSM and presented the ground state (g.s) and the excited $2^{+}_{1}$ state energies, the B(E2; $2^{+}_{1} \rightarrow 0^{+}_{1}$) transition rates and the $2^{+}_{1}$ quadrupole moments dependence on the NCSM parameters. 

The elastic electron scattering and muonic atom x-ray spectroscopy are employed to measure the charge radius of the nucleus. Since these techniques are applied only for stable isotopes, therefore, a different method is required for radii measurement of neutron-rich isotopes. The optical and $K_{\alpha}$ x-ray isotopes shift (IS) method is the only way to determine the charge radii of short-lived unstable nuclei, although it is also challenging to apply this method for the nuclei lying in the range $4<Z<10$ due to inadequate precision in the atomic physics calculations and difficulty in low-energy isotopes production. In Ref. \cite{Kanungo}, a new technique, \textit{viz.} charge changing cross-section ($\sigma_{CC}$) measurement, has been used in probing the point-proton radii ($r_p$) of $^{12-19}$C isotopes. The measured $r_p$ for $^{18}$C and $^{19}$C was 2.39(4) and 2.40(3) fm, respectively. The $r_p$ of $^{18}$C was in good agreement with those obtained from \textit{ab initio} coupled-cluster theory using chiral \textit{NN} and \textit{NNN} interactions (NNLO$_{\rm sat}$). Also, $r_p$ for $^{12-16}$C isotopes were extracted using $\sigma_{CC}$ technique at the Research Center for Nuclear Physics (RCNP) facility, Osaka, and the Glauber model within the optical-limit approximation \cite{Tran1}. In Ref. \cite{Tran2}, the smallest spin-orbit originated magic number at $Z=6$ in $^{12-20}$C isotopes was evident from the $r_p$ distribution and B(E2) measurements. The $r_p$ for $^{12-19}$C was observed to be almost constant, which might be an indication of an inert proton core.

In the present paper, we discuss the evolution of nuclear structure for $^{18-20}$C  isotopes by using the \textit{ab initio} NCSM method with realistic \textit{NN} interactions. We have calculated low-lying energy spectra of $^{18-20}$C isotopes and compared them with the experimental data. In addition, electromagnetic observables of these isotopes are predicted and compared with the experimental data wherever available. We have reported proton and neutron occupancies for the ground and first excited states with each interaction. Also, point-proton radii of g.s. of $^{18-20}$C are evaluated and their variations with NCSM parameters are shown. Thus, our work is a comprehensive study of $^{18-20}$C isotopes using the NCSM, with $^{19}$C being investigated for the first time.

This paper is divided into the following sections. In Section II, the NCSM formalism is briefly summarized. In Section III, we introduce the realistic \textit{NN} interactions. Further, we present the calculated results of energy spectra, spectroscopic properties, occupancies and point-proton radii of $^{18,19,20}$C isotopes obtained from the NCSM approach in Section IV. Finally, we conclude the paper in Section V. 

\section{\textit{Ab initio} no-core shell model formalism}
We consider a nuclear system that consists of $A$ point like non-relativistic nucleons interacting via realistic interaction. We have constrained our calculations up to two-body interactions. In NCSM approach \cite{2009,nocore_review}, these $A$ nucleons of the system are taken as the degree of freedom. The initial Hamiltonian for this system is written as
\begin{equation}\label{Eq.(1)}
		H_{A}  = \frac{1}{A}\sum_{i<j}^{A} \frac{{(\vec p_i - \vec p_j)}^2}{2m} + \sum_{i<j}^{A} V^{NN}_{ij},
\end{equation}
where, the first term indicates the relative kinetic energy operator, and the second term is the $NN$ interaction containing both nuclear and Coulomb parts. Here, $m$ is the mass of nucleon. In the NCSM method, harmonic oscillator (HO) basis states are employed that are restricted by truncation parameter N$_{\mathrm{max}}$. N$_{\mathrm{max}}$ is defined as the maximum number of allowed HO excitations above the g.s. configuration of $A$ nucleons system. Because of the properties of these realistic nuclear interactions, we require effective \textit{NN} interaction in order to obtain convergent results. We add centre-of-mass (c.m.) HO Hamiltonian to 
the starting Hamiltonian \ref{Eq.(1)} to facilitate the derivation of effective Hamiltonian. Now, modified Hamiltonian will take the form
\begin{equation}\label{Eq.(2)}
	\begin{split}
	H^{\Omega}_{A} & = H_{A} + H_{\mbox{c.m.}} 
	    = \sum_{i=1}^{A} \left[\frac{{\vec p_i}^2}{2m}+\frac{1}{2}m {\Omega}^2 {\vec r_i}^2 \right] \\
	&   + \sum_{i<j}^{A} \left[V_{ij}^{NN}-\frac{m {\Omega}^2}{2A} {(\vec r_i - \vec r_j)}^2 \right], 
	   \end{split}
\end{equation}
where, the first term indicates a sum of single particle HO Hamiltonian. Here, the c.m. HO Hamiltonian is 
$		H_{\mbox{c.m.}} = \frac{\vec{P}^2}{2mA} + \frac{1}{2}Am{\Omega}^2{\vec R}^2,$
	  \text{where}, $\Omega$ represents the HO frequency, 
$	 \vec{P} = \sum_{i=1}^{A}\vec p_i$
and
$
	\vec R = \frac{1}{A} \sum _{i=1}^{A} \vec r_i.
$
The Hamiltonian \ref{Eq.(2)} can be rewritten as
\begin{equation*}
	H^{\Omega}_{A} = \sum _{I=1}^{A} h_{i} + \sum _{i<j}^{A}V^{\Omega,A}_{ij}.
\end{equation*}
Hence, the modified Hamiltonian becomes frequency dependent. This modified Hamiltonian has the same intrinsic eigenstates as obtained from the translationally invariant Hamiltonian \ref{Eq.(1)}. To solve the Schr{\"o}dinger equation of $A$ nucleons system, the full Hilbert space is split into two parts which are finite model space ($P$) that contains all HO basis states up to N$_{\mathrm{max}}$ and the remaining model space $Q$  ($= 1-P$). Thus, final NCSM calculations are performed in the $P$ model space. To derive effective Hamiltonian, two renormalization procedures are implemented that are the Okubo-Lee-Suzuki (OLS) scheme \cite{Prog.Theor.Phys.12,Prog.Theor.Phys.,Prog.Theor.Phys.68,Prog.Theor.Phys.92} and the Similarity Renormalization Group (SRG) \cite{SRG1,SRG2}. For the interaction used in our calculations, former unitary transformation is employed. Now, we replace the second term in the Hamiltonian \ref{Eq.(2)} by the effective interaction and  subtract out the c.m. Hamiltonian, which is added prior. So, the effective Hamiltonian will have the form
\begin{equation}\label{Eq.(3)}
\begin{split}
	H_{A,\rm eff}^{\Omega}  &=  P\left\{ \sum _{i<j}^{A} \left[ \frac{{(\vec p_i - \vec p_j)}^2}{2mA} 
	   + \frac{m {\Omega}^2}{2A} {(\vec r_i - \vec r_j)}^2 \right]\right. \\
&	\left.+\sum_{i<j}^{A} \left[ V^{NN}_{ij} - \frac{m {\Omega}^2}{2A}{(\vec r_i - \vec r_j)}^2\right]_{\rm eff} \right\}P.
	\quad 
\end{split}
\end{equation}
This effective Hamiltonian has up to $A$-body terms and it is rather difficult to solve this $A$-body system, thus, a simple approximation which is a two-body cluster OLS is used. Since Hamiltonian \ref{Eq.(2)} is not translationally invariant, it generates spurious eigenstates with excited c.m. motion. These spurious states are projected upward in the energy spectrum by the inclusion of the Lawson projection term $\beta \Big(H_{\mbox{c.m.}} - \frac{3}{2}\hbar\Omega\Big)$ \cite{Lawson} to the Hamiltonian \ref{Eq.(3)}, in the final step. In our calculation, $\beta$ is taken 10 that is  large enough to shift these spurious states up to high energies. This term will not change the intrinsic eigenstates of the system. In this way, our final calculations will vary with HO frequency and N$_{\mathrm{max}}$.

\section{Details about the effective \textit{NN} interactions}
In the present NCSM calculations, the CDB2K~\cite{Machleidt1,Machleidt2,Machleidt3,Machleidt4}, INOY~\cite{INOY,nonlocal,Doleschall} and N\textsuperscript{3}LO~\cite{QCD,Machleidt}  realistic \textit{NN} interactions have been  adopted. These interactions  come from either meson-exchange theory or Quantum Chromodynamics (QCD) via chiral effective field theory ($\chi$EFT). 
In the CDB2K  interaction, exchange particles between two nucleons are mesons with masses below the nucleon mass. These mesons are pions ($\pi^{\pm,0}$), rho ($\rho^{\pm,0}$), $\eta$, $\omega$ \textit{etc}. 
The INOY interaction contains local character at long distance while it shows non-local behavior at short distance ($r < 3$ fm) and the form of this potential can be found in Refs. ~\cite{INOY,nonlocal}. Its non-local feature helps to reproduce correct binding energies of $^3$H and $^3$He without the addition of three-body forces. 
 The Bonn-Bochum-J\"{u}lich group has developed the chiral \textit{NN} potentials, which are reported in Refs.~\cite{Epelbaum1,Epelbaum2,Epelbaum3,Epelbaum4,Epelbaum5}.
The N\textsuperscript{3}LO interaction, firstly presented by the Idaho group, Entem and Machleidt \cite{QCD,Machleidt}, is based on the expansion up to fourth order of chiral perturbation theory.
The CDB2K, INOY and N\textsuperscript{3}LO interactions present strong short range correlations, therefore, require the OLS renormalization. NCSM calculations for these \textit{ab initio} interactions are performed using pAntoine \cite{pAntoine1,pAntoine11,pAntoine2}.

Additionally, the shell model calculations with the phenomenological YSOX interaction \cite{YSOX} have been carried out for comparison with the \textit{ab initio} results. In YSOX interaction, $^{4}$He is taken as the inert core and it contains $psd$-model space. We have performed calculations in a $psd$-model space with 2$\hbar\Omega$ excitations using KSHELL code \cite{KSHELL2019}.

As reported in \cite{Navratil2007}, the $NNN$ force improves the excitation energy spectra of light nuclei. However, it is also mentioned that the $NNN$ force over-corrects the deficiencies of the $NN$ interaction in the case of the $^{12}$C and shifts the $7/2^-$ state of $^{13}$C upwards in the energy spectrum, which indicates that further improvement in the interaction is required. Also, the three-nucleon force leads to an overbinding of the carbon isotopes, as reported in Ref.~\cite{Maris2021}; this issue is still unresolved as to whether it is caused due to deficiencies of the N$^2$LO approximation to the $NN$ interaction, or, has to be fixed by higher-order $NNN$ force contributions. 
Earlier, our group have successfully studied boron ($^{10-14}$B) \cite{pc1}, nitrogen ($^{18-22}$N) \cite{arch1}, oxygen ($^{18-23}$O) \cite{arch2} and florin ($^{18-24}$F) \cite{arch2} isotopes using NCSM method and we found that without inclusion of the three-body forces, INOY interaction correctly reproduces g.s. spin-parity of $^{10}$B as $3^+$ and also predicts the correct location of drip-line for oxygen chain at $^{24}$O.
Since inclusion of nonlocality in the INOY $NN$ interaction can account some of the many-nucleon force effects, thus, our aim in the present work is to check whether or not this interaction improves the description of near drip line carbon isotopes. Also, we want to test which interaction is more suitable for these nuclei. So, we have used three different realistic $NN$ interactions.

\begin{figure*}
	\includegraphics[width=6cm]{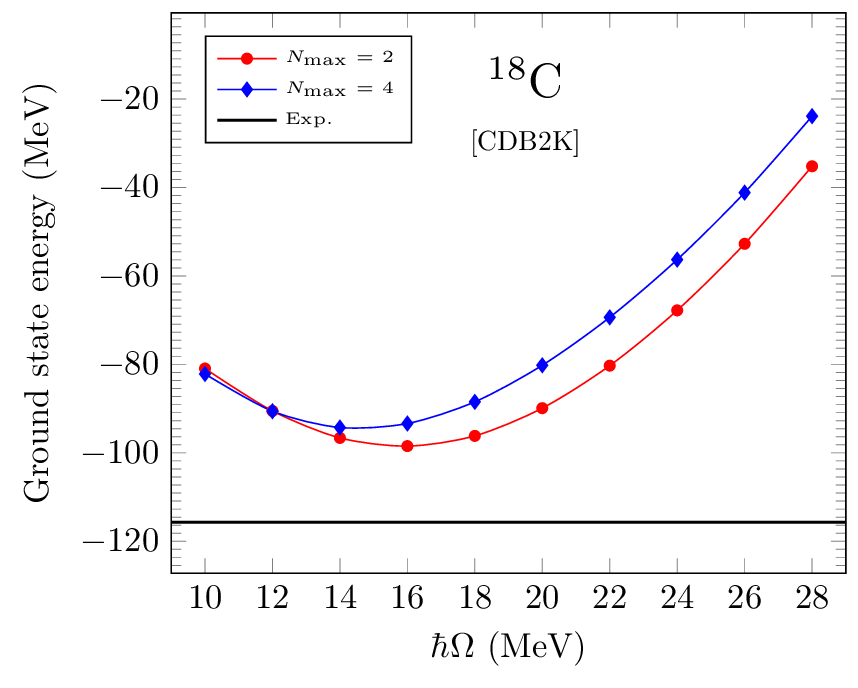}
	\includegraphics[width=6cm]{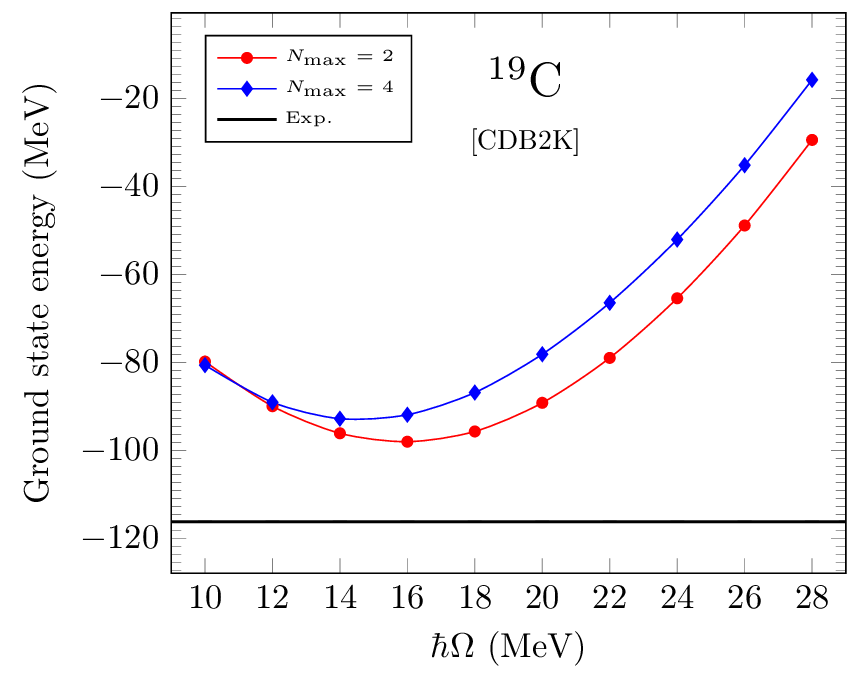}
	\includegraphics[width=6cm]{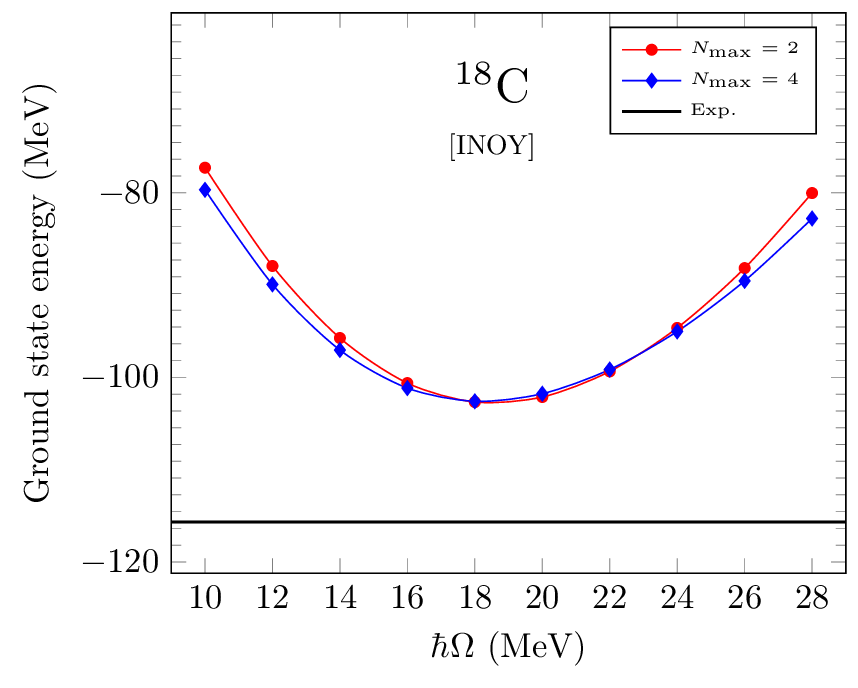}
	\includegraphics[width=6cm]{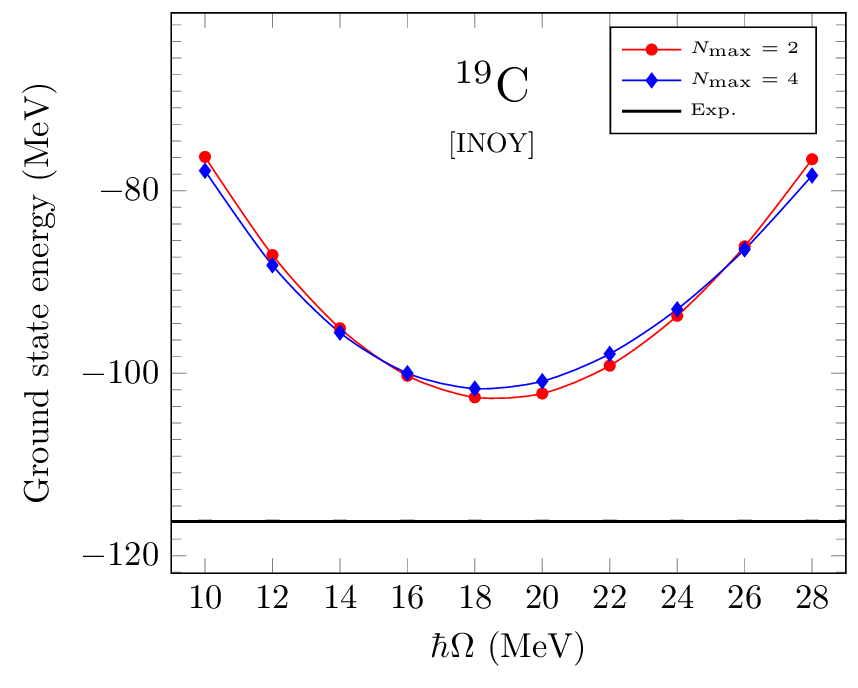}
	\includegraphics[width=6cm]{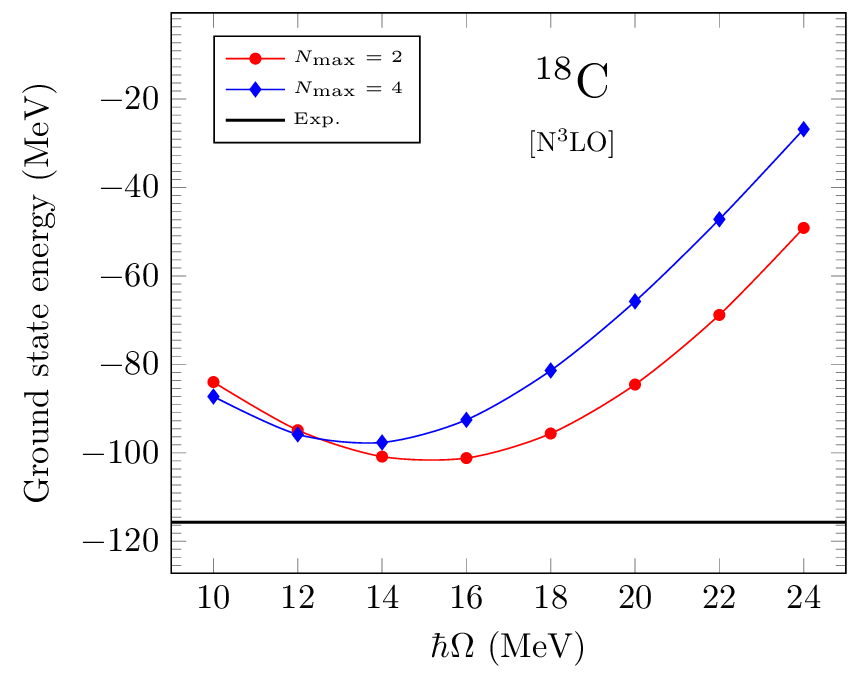}
	\includegraphics[width=6cm]{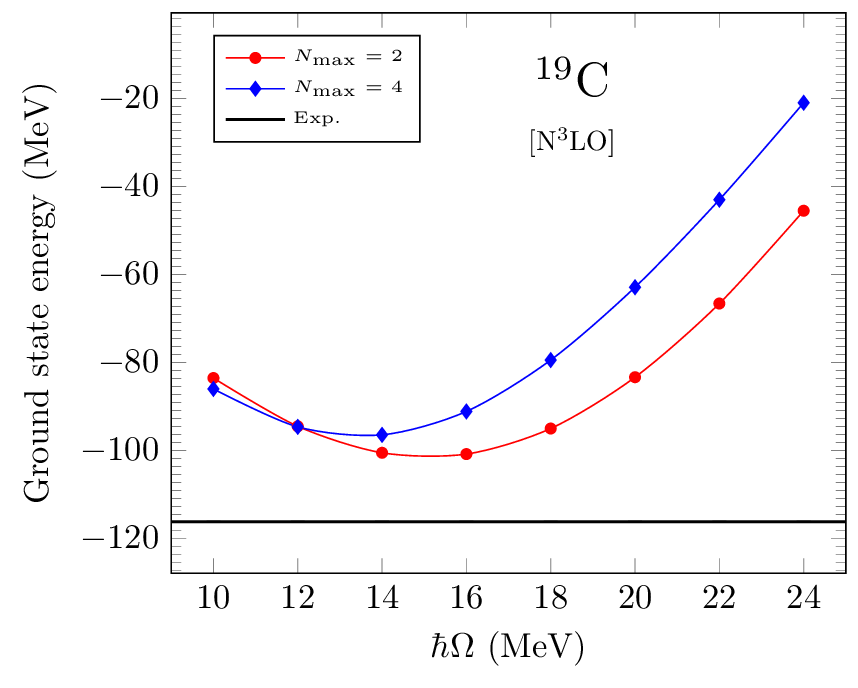}
	\caption{ Dependence of ground state energy for $^{18}$C and $^{19}$C on HO frequency at different $N_{\mathrm{max}}$ corresponding to the CDB2K, INOY and N\textsuperscript{3}LO interactions. Horizontal line with error bar represents the experimental binding energy.}
	\label{GS_CDB2K_INOY}
\end{figure*}

\begin{figure*}
	\includegraphics[width=6cm]{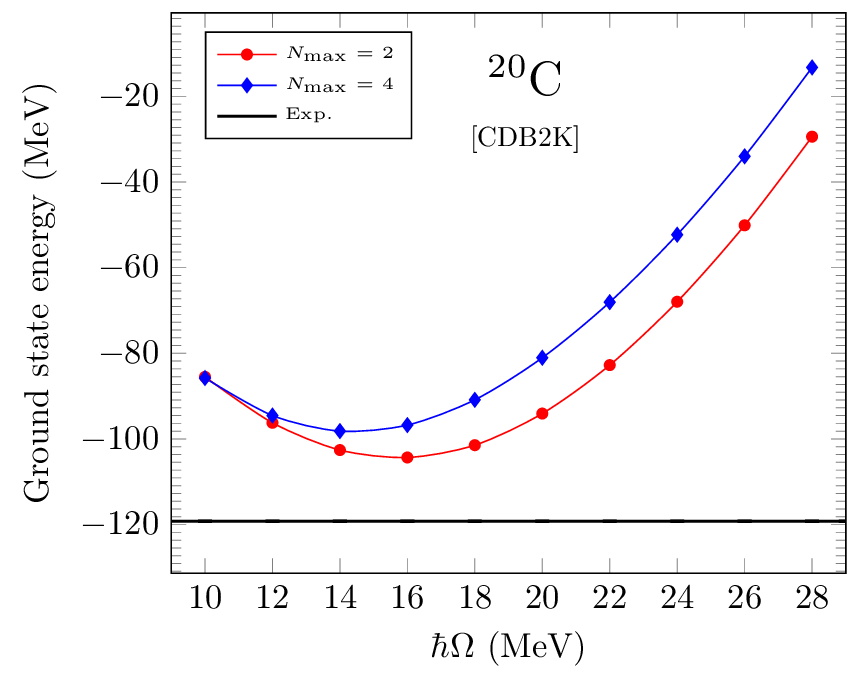}
	\includegraphics[width=6cm]{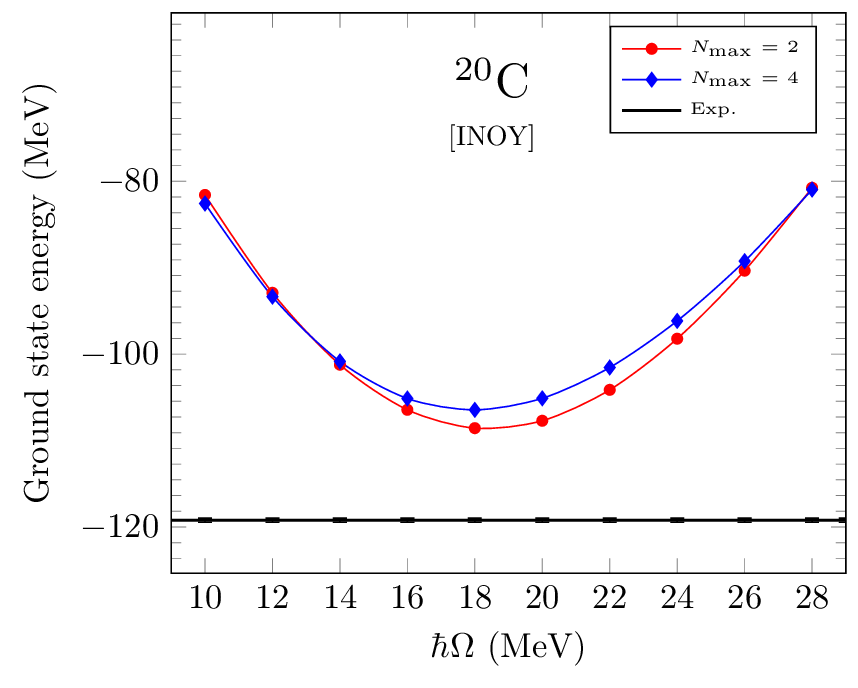}
	\begin{center}
		\includegraphics[width=6cm]{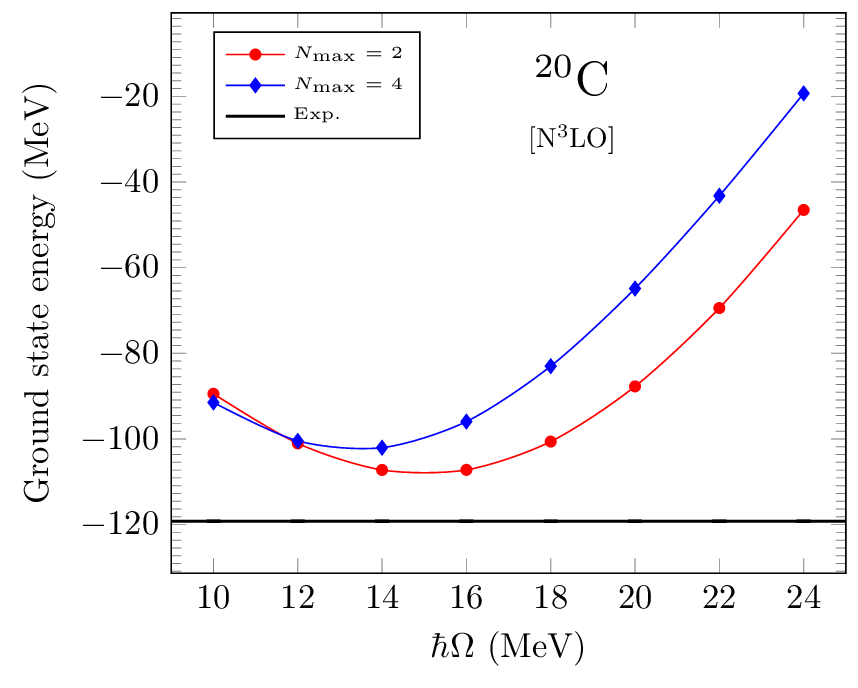}
	\end{center}
	\caption{ Dependence of ground state energy for $^{20}$C on  HO frequency at different$N_{\mathrm{max}}$ corresponding to the CDB2K, INOY and N\textsuperscript{3}LO interactions. Horizontal line with error bar represents the experimental binding energy.}
	\label{GS_N3LO_N2LOopt}
\end{figure*}
\section{Results and Discussion}
It is well known that the dimension of the Hamiltonian matrix drastically increases with increasing basis size and neutron number across an isotopic chain. We are able to perform the NCSM calculations up to N$_{\mathrm{max}}$ = 4 for $^{18}$C, $^{19}$C and $^{20}$C isotopes, and the corresponding m-scheme dimensions are $5.2\times10^{7}$, $9.0\times10^{7}$ and $1.3\times10^{8}$, respectively. Importance Truncation NCSM (IT-NCSM) is a modified NCSM approach that is successfully used to tackle large matrix diagonalization. Using IT-NCSM approach, Fors{\'s}en \textit{et al.} \cite{Forssen} have reached up to N$_{\mathrm{max}}$ = 8 and N$_{\mathrm{max}}$ = 6 for $^{18}$C and $^{20}$C, respectively. 

Firstly, optimal frequency is selected for each isotope corresponding to different interactions in the NCSM method. In Figs. \ref{GS_CDB2K_INOY}-\ref{GS_N3LO_N2LOopt}, we have shown the variation of binding energy with HO frequencies and basis sizes. The N$_{\mathrm{max}}$-dependent binding energy curves indicate a decrease in dependence of binding energy on HO frequency with increasing N$_{\mathrm{max}}$. The optimal frequency corresponds to the minimal value of g.s. energy for largest N$_{\mathrm{max}}$. We found that optimal frequency for $^{18}$C is 14, 18 and 14  MeV for CDB2K, INOY  and N$^{3}$LO, respectively. In a similar way, we obtain optimal frequencies for other carbon isotopes corresponding to different interactions. We note that the same optimal frequency is obtained for $^{18}$C, $^{19}$C and $^{20}$C in the case of CDB2K. This is also the case for INOY and N$^{3}$LO interactions. We have studied near drip line carbon nuclei with natural parity states, therefore, only even values of N$_{\mathrm{max}}$ are taken into account. 
\begin{figure*}
	\includegraphics[width=\textwidth]{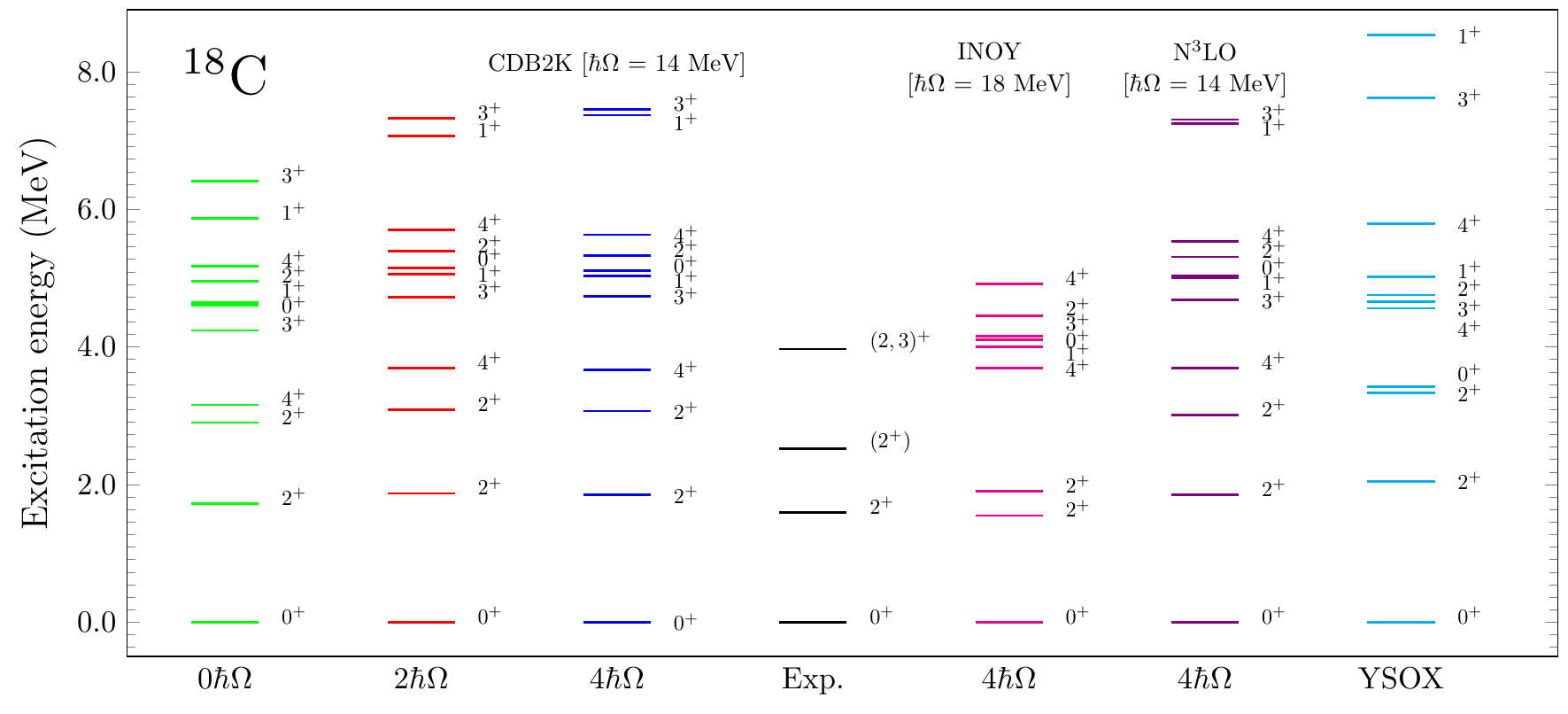}
	\includegraphics[width=\textwidth]{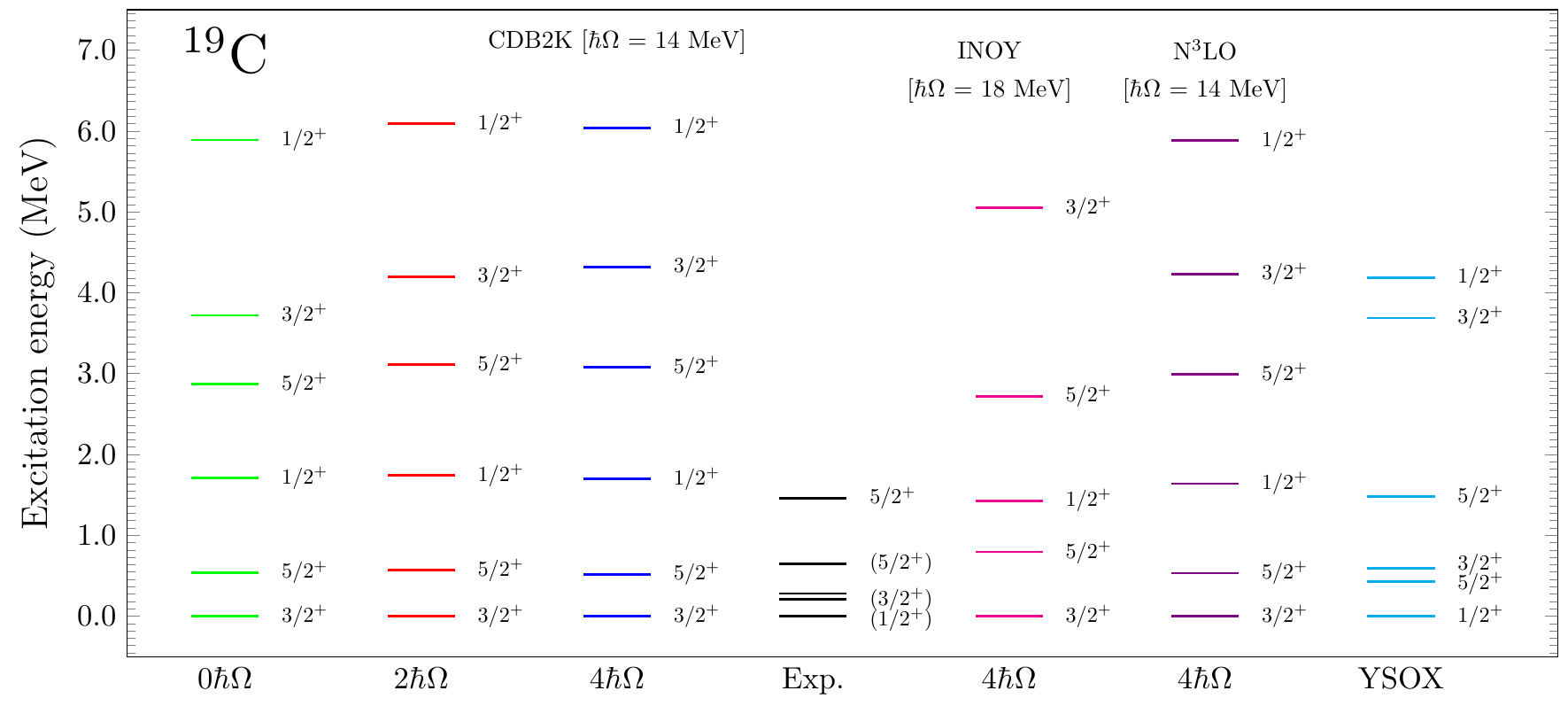}
	\includegraphics[width=\textwidth]{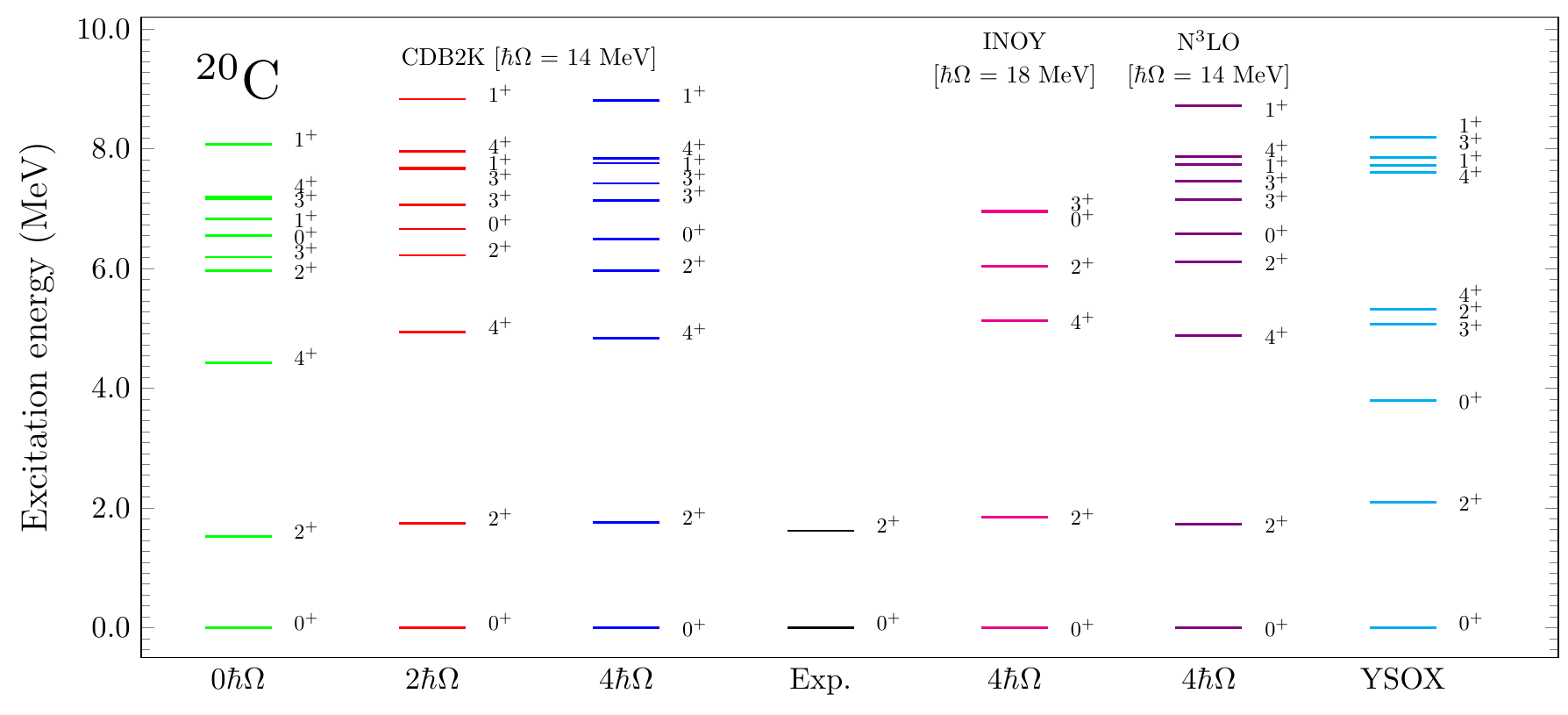}
	\caption{Low-lying energy spectra of $^{18-20}$C isotopes using \textit{ab initio} as well as phenomenological interactions.}
	\label{Spectra}
\end{figure*}
In Fig. \ref{Spectra}, we have presented the energy spectra of $^{18-20}$C isotopes using three realistic interactions. In case of the CDB2K interaction, convergence behavior of excited states are shown from $N_{\mathrm{max}}$ = 0 to $N_{\mathrm{max}}$ = 4. For other interactions, the results corresponding to the largest model space are shown.  Experimental data are taken from Ref. \cite{NNDC}. For comparison, results corresponding to the YSOX interaction are also reported in the last column. At the top of each figure, interactions with their respective optimal frequencies are mentioned.

For $^{18}$C, g.s. spin-parity is well reproduced with all three microscopic interactions. As for the excited states, the spin-parity of only the first excited state is confirmed, which is $2^+$, and the excitation energy of this state is 1.588(8) MeV. From \textit{ab initio} calculations, the excitation energies for the above mentioned state are 1.855, 1.550  and 1.849 MeV using the CDB2K, INOY and  N$^{3}$LO, respectively.  With the YSOX interaction, the energy of this state comes out to be slightly higher compared to \textit{ab initio} interactions. Using all interactions, the spin and parity of second excited state is confirmed to be $2^+$, although experimentally it is tentative. The INOY interaction underpredicts the excitation energy of $2_2^+$, while the CDB2K and N$^{3}$LO interactions overpredict.   Experimentally, a state at 3.972(20) MeV could be either $2^+$ or $3^+$, therefore, we have calculated both these spin states and found that the energy of $3^+$ state is lower than $2^+$. Hence, on the basis of energy, $3^+$ is the more preferred spin for this state.  We have also calculated some excited states up to spin 4. 
In case of $^{19}$C, measured g.s. spin is not confirmed yet, and it is assigned as $(1/2^+)$. None of the \textit{ab initio} interactions is able to reproduce the experimental g.s. of the same. The CDB2K, INOY, and N$^{3}$LO interactions predict the g.s. as $3/2^+$. The calculated first excited state, $5/2^+$, using CDB2K, INOY, and N$^{3}$LO is in reasonable agreement with the experimental value, although experiment gives two more excited states below this state. YSOX interaction predicts the g.s. as $1/2^+$, similar to the experimental result, and first excited state to be $5/2^+$ as obtained with CDB2K, INOY,  and N$^{3}$LO interactions.
For $^{20}$C, only experimental g.s. and the first excited state are known with spins $0^+$ and $2^+$, respectively. \textit{Ab initio} as well as YSOX interactions reproduce spins of these states correctly. The energy separation between these states is obtained as 1.757, 1.848  and 1.730  MeV from the CDB2K, INOY  and N$^{3}$LO, respectively, which is close to the experimental value of 1.618(11) MeV. Similar to $^{18}$C, we have also predicted some excited states with spins in the range 0 to 4. 
\begin{table*}
	\centering
	\caption{\label{gsenergy} Comparison of experimental ground state energy  with \textit{ab initio} results in largest basis size and YSOX results for $^{18-20}$C isotopes.}
	\centering
	\begin{tabular}{cccccc}
		\vspace{-2.8mm}\\
		\hline
		\hline	
			\vspace{-2.8mm}\\
		Nucleus & Exp. & CDB2K & INOY  & N$^3$LO  & YSOX \\
		\hline 	\vspace{-2.8mm}\\
		$^{18}$C&-115.670(31) & -94.275 &-102.593 &-97.657   & -115.670 \\
		$^{19}$C& -116.242(95) &  -92.788& -101.687& -96.459 & -115.410 \\
		$^{20}$C&-119.22(24) &  -98.169& -106.447& -102.042& -119.231 \\
		\hline\hline
	\end{tabular}	
\end{table*}

Calculated g.s. energy of $^{18-20}$C isotopes is presented in Table \ref{gsenergy} corresponding to the CDB2K, INOY and N$^{3}$LO  interactions. One can say that the energy obtained from the INOY interaction is closest to the experimental data among all the \textit{ab initio} interactions, still, the deviation from the experimental data is $\sim$ 11\%. For further improvement in the g.s. energy, one needs to enlarge the basis size beyond N$_{\mathrm{max}}$ = 4. We observe from Table \ref{gsenergy} that all the microscopic interactions underbind the g.s. of these isotopes.

We have calculated electromagnetic observables such as B(E2), electric quadr-upole moment (Q), reduced magnetic dipole transition strength B(M1), and magnetic dipole moment ($\mu$) for $^{18-20}$C isotopes, which are presented in Table \ref{EM_properties}. These electromagnetic observables are reported corresponding to the largest $N_{\rm max}$ and optimal frequency. For $^{18}$C, B(E2) value for the transition from first $2^+$ to the g.s. was experimentally studied from a lifetime measurement and its value was deduced as 4.3 $\pm$ 1.2 $\rm e^2 \rm fm^4$ (1.5 $\pm$ 0.5 W.u.)  \cite{Ong}. The $^{18}$C is considered as an open-shell nucleus, still, its measured B(E2) value was even more suppressed than those of the doubly magic nuclei. Further, $^{18}$C was populated by the one-proton knockout reaction of $^{19}$N, and the extracted B(E2) value was $3.64\substack{+0.15\\-0.14}(\rm stat)\substack{+0.40\\-0.47}(\rm syst)$ $\rm e^2 \rm fm^4$ \cite{18C_2012}. Also, authors have compared measured results with the \textit{ab initio} NCSM calculations using CDB2K and INOY interactions and the corresponding values were obtained as 4.2(4) and 1.5 $\rm e^2 \rm fm^4$, respectively. In our calculations, we get B(E2) = 3.134 $\rm e^2 \rm fm^4$ from CDB2K interaction, while INOY interaction underpredicts it. The calculated value using N$^3$LO interaction is 3.364 $\rm e^2 \rm fm^4$, which is in a reasonable agreement with the experimental data. In the case of YSOX, we have performed calculations with two sets of effective charges. With standard effective charges (set I) $e_p = 1.5e$ and $e_n = 0.5e$, we get quite large B(E2) value [9.600 $\rm e^2 \rm fm^4$]. In another case (set II), $e_p = 1.11e$ and $e_n = 0.27e$ are used as suggested in Ref. \cite{20C_2011} and corresponding B(E2) value for $2^+$ to $0^+$ transition is 3.490 $\rm e^2 \rm fm^4$ which supports the experimental value \cite{18C_2012}. We have also given predictions for B(E2; $2_2^+ \rightarrow 0_1^+$) to guide future experiments. For $^{19}$C, our calculated B(E2; $5/2_1^+ \rightarrow 1/2_1^+$) values are 0.008, 0.014,  and 0.003 with CDB2K, INOY,  and N$^{3}$LO  interactions, respectively. We have adopted effective charge $e_p = 1.09e$, $e_n = 0.24e$ \cite{SHF} for Set II calculations in $^{19}$C. 
\begin{sidewaystable}
	\centering
	\caption{\label{EM_properties} Calculated B(E2), $Q$, B(M1) and $\mu$ for $^{18-20}$C isotopes using three realistic interactions as well as YSOX. For \textit{ab initio} calculations, bare charges $e_p = 1.0e$, $e_n = 0.0e$ and g$_s^\mathrm{eff}$ = g$_s^\mathrm{free}$ are used; while for YSOX calculations, different effective charges,  g$_{l}$, and g$_{s}$ are used as shown in the footnote.}
		\begin{tabular}{cccccccc}
			\vspace{-2.8mm}\\
			\hline \hline
			&        & \multicolumn{6}{c}{~~~~~~~~~~~~~~~~\hspace{5cm}B(E2) ($e^2fm^4$)}   \T\B \\
			\cline{3-8}
			Nuclei  & Transition & Exp. & CDB2K  & INOY  & N$^3$LO & YSOX(Set I)\footnote{\label{bare}$e_p = 1.5e$ and $e_n = 0.5e$} & YSOX (Set II)\T\B	\\
			\hline
			$^{18}$C      &	$2_{1}^{+}$$\rightarrow$$0_{1}^{+}$  & $4.3\pm1.2$ \cite{Ong}, $3.64\substack{+0.15\\-0.14}(\rm stat)\substack{+0.40\\-0.47}(\rm syst)$ \cite{18C_2012}   & 3.134     & 1.507     & 3.364       & 9.600&3.490\footnote{\label{eff}$e_p = 1.11e$ and $e_n = 0.27e$ \cite{SHF,20C_2011}}  \\
			&	$2_{2}^{+}$$\rightarrow$$0_{1}^{+}$  &  NA   & 0.070     & 0.029     & 0.058       & 2.800 &1.229\textsuperscript{\ref{eff}} \T\B\\
			$^{19}$C      &		$5/2_1^+$$\rightarrow$$1/2_1^+$  &  NA   & 0.008     & 0.014     & 0.003       & 0.700 & 0.302\footnote{\label{19eff}$e_p = 1.09e$ and $e_n = 0.24e$ \cite{SHF}}\\
			$^{20}$C      &	$2_{1}^{+}$$\rightarrow$$0_{1}^{+}$  & $<$3.68 \cite{20C_2009}, $7.5\substack{+3.0\\-1.7}(\rm stat)\substack{+1.0\\-0.4}(\rm syst)$ \cite{20C_2011}  & 3.865     & 2.099     & 4.080     & 15.700 & 4.901\footnote{\label{20eff}$e_p = 1.07e$ and $e_n = 0.22e$ \cite{SHF,20C_2011}}\\
			\hline
			&        & \multicolumn{6}{c}{~~~~~~~~~~~~~~~~\hspace{5cm}Q (eb)}   \T\B  \\
			\cline{3-8}
			Nuclei  & J$^{\pi}$ & Exp. & CDB2K  & INOY  & N$^3$LO  & YSOX(Set I)\textsuperscript{\ref{bare}} & YSOX (Set II)\T\B	\\
			\hline
			$^{18}$C      &		$2_1^+$  &  NA   & 0.035     & 0.011     & 0.036      &  -0.053&-0.031\textsuperscript{\ref{eff}}\\
			$^{19}$C      &		$3/2_1^+$  &  NA   & -0.026     & -0.018     & -0.027       & -0.050& -0.028\textsuperscript{\ref{19eff}}\\
			$^{20}$C      &		$2_1^+$  &  NA   & 0.039     & 0.027     & 0.040      &  0.079& 0.044\textsuperscript{\ref{20eff}}\T\B\\
			\hline			&        & \multicolumn{6}{c}{~~~~~~~~~~~~~~~~\hspace{5cm}B(M1) ($\mu_N^2$)}  \T\B   \\
			\cline{3-8}
			Nuclei  & Transition & Exp. & CDB2K  & INOY  & N$^3$LO  & YSOX(Set I)\footnote{\label{set1}g$_{lp}^\mathrm{eff}$ = 1.0, g$_{ln}^\mathrm{eff}$ = 0.0, g$_{sp}^\mathrm{eff}$ = 5.585, and g$_{sn}^\mathrm{eff}$ = -3.826} & YSOX (Set II)\footnote{\label{set2}g$_{lp}^\mathrm{eff}$ = 1.175, g$_{ln}^\mathrm{eff}$ = -0.106, g$_{sp}^\mathrm{eff}$ = 5 and g$_{sn}^\mathrm{eff}$ = -3.5 \cite{19C_2015}}\T\B	\\
			\hline
			$^{18}$C      &	$1_{1}^{+}$$\rightarrow$$0_{1}^{+}$ &  NA   & 0.021    & 0.001     & 0.023      & 0.001 & 0.001\\
			$^{19}$C      &		$3/2_{1}^{+}$$\rightarrow$$1/2_1^+$  &  0.00321(25) \cite{19C_2015}   & $<$0.001    & 0.003     & $<$0.001       & $<$0.001 & $<$0.001\\
			&		 $5/2_{1}^{+}$$\rightarrow$$3/2_1^+$  &  NA  & 0.190     & 0.064     & 0.199     & 0.022 &0.036\\
			$^{20}$C      &		$1_{1}^{+}$$\rightarrow$$0_{1}^{+}$& NA   & 0.253     & 0.121     & 0.233      & 0.023 & 0.022\\
			\hline
			&        & \multicolumn{6}{c}{~~~~~~~~~~~~~~~~\hspace{5cm}$\mu$ ($\mu_N$)} \T\B  \\
			\cline{3-8}
			Nuclei  & J$^{\pi}$ & Exp. & CDB2K  & INOY  & N$^3$LO  & YSOX(Set I)\textsuperscript{\ref{set1}}& YSOX (Set II)\textsuperscript{\ref{set2}} \T\B\\
			\hline
			$^{18}$C      &		$2_1^+$  &  NA   & -0.455     & -0.466     & -0.373      &  -0.202 & -0.353\\
			$^{19}$C      &		$1/2_1^+$  &  NA   & -1.530    & -1.396     & -1.548       & -1.006 &-1.330\\
			&		$3/2_1^+$  &  NA   & 0.531     & 0.194     & 0.563      & 0.226&0.131 \\
			$^{20}$C      &		$2_1^+$  &  NA   & 0.148     & -0.015     & 0.186      &  -0.072 & -0.186\\		    		    			
			\hline
			\hline
		\end{tabular}
	\label{Moments}
\end{sidewaystable}
In the case of near dripline nucleus $^{20}$C, the measured B(E2) value for the  $2_1^+ \rightarrow 0_{\rm g.s.}^+$ transition is $7.5\substack{+3.0\\-1.7}(\rm stat)\substack{+1.0\\-0.4}(\rm syst)$ $\rm e^2 \rm fm^4$ \cite{20C_2011}, this was obtained from the lifetime measurement of the $2_1^+$ state at NSCL-MSU. In this work, the shell model calculation in $psd$-model space using WBT interaction \cite{Warburton} was also carried out and value of B(E2) was obtained as 6.80 $\rm e^2 \rm fm^4$ with $e_p = 1.07e$ and $e_n = 0.22e$. The suppressed value of effective charges can be a signature of weak binding and decoupling of valence neutrons from the core. Since effective charges have $1/A$ dependence, these effective charges are normal in this region. Thus, $^{20}$C is not a candidate for a halo structure. Earlier \cite{20C_2009}, from an inelastic scattering measurement, the B(E2;$2_1^+ \rightarrow 0_{\rm g.s.}^+$) was derived as $< 3.68 \,\, \rm e^2 \rm fm^4$. These two different experimentally measured B(E2) values emerge as an interesting case for us to  perform \textit{ab initio} calculations. In our calculations, chiral interaction provides B(E2) value of around 4 $\rm e^2 \rm fm^4$ which is consistent with Ref. \cite{20C_2009}. The B(E2) value calculated with the INOY interaction is small as compared to the experimental value \cite{20C_2009}. 
In Ref. \cite{20C_2011}, the shell model calculation predicts that the shell gap between $p_{1/2}$ and $p_{3/2}$ decreases from $^{16}$C to $^{22}$C. As a consequence, a large value of B(E2) is obtained for $^{20}$C, which can be attributed to proton admixture to the $2_1^+$ state.  Similar conclusions are reported in Ref. \cite{neutronrichC}. Our \textit {ab initio} results do not support this finding (such large B(E2) value).

\begin{figure*}
	\centering
	\includegraphics[width=8.0cm]{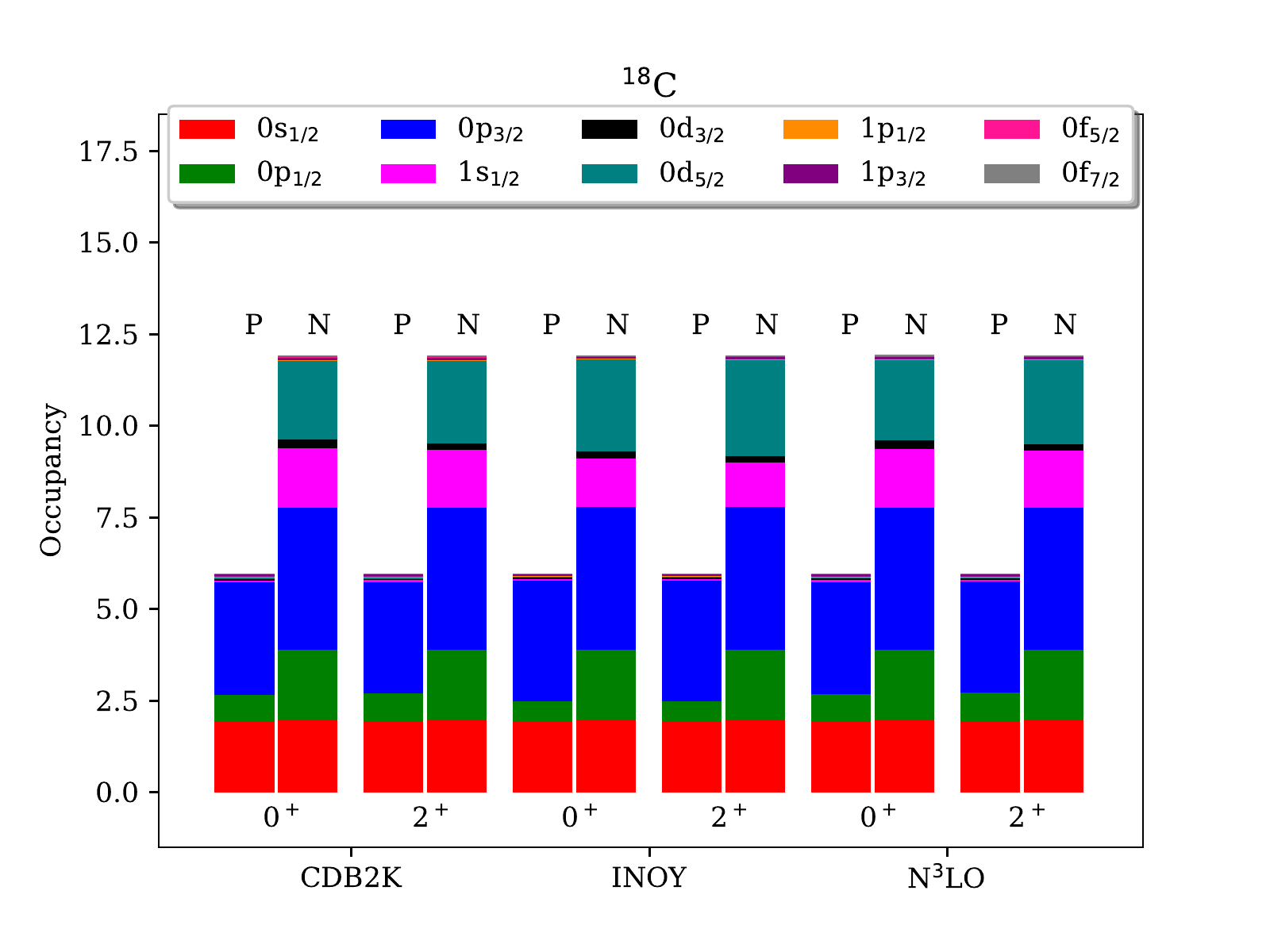}
	\includegraphics[width=8.0cm]{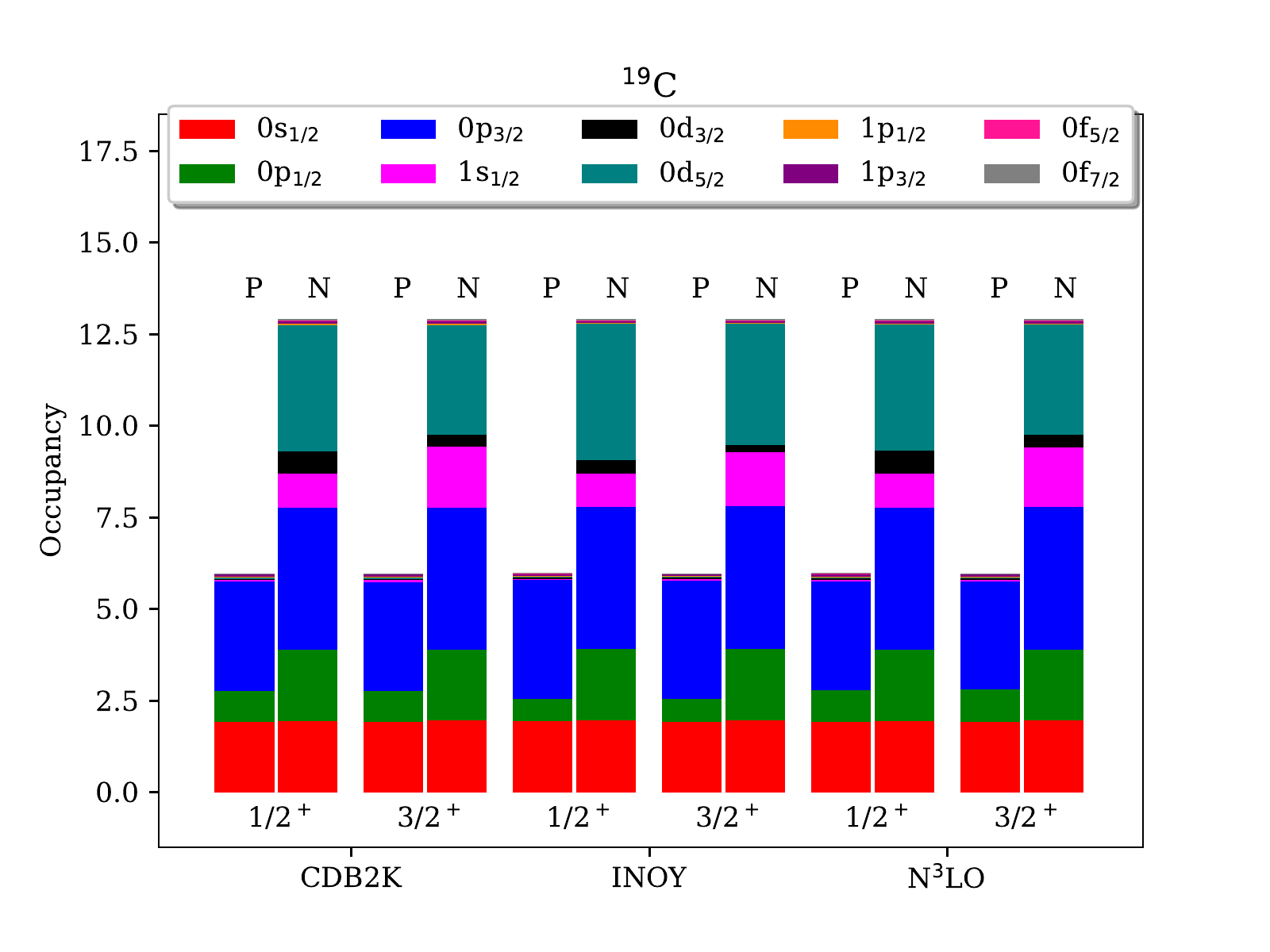}
	\includegraphics[width=8.0cm]{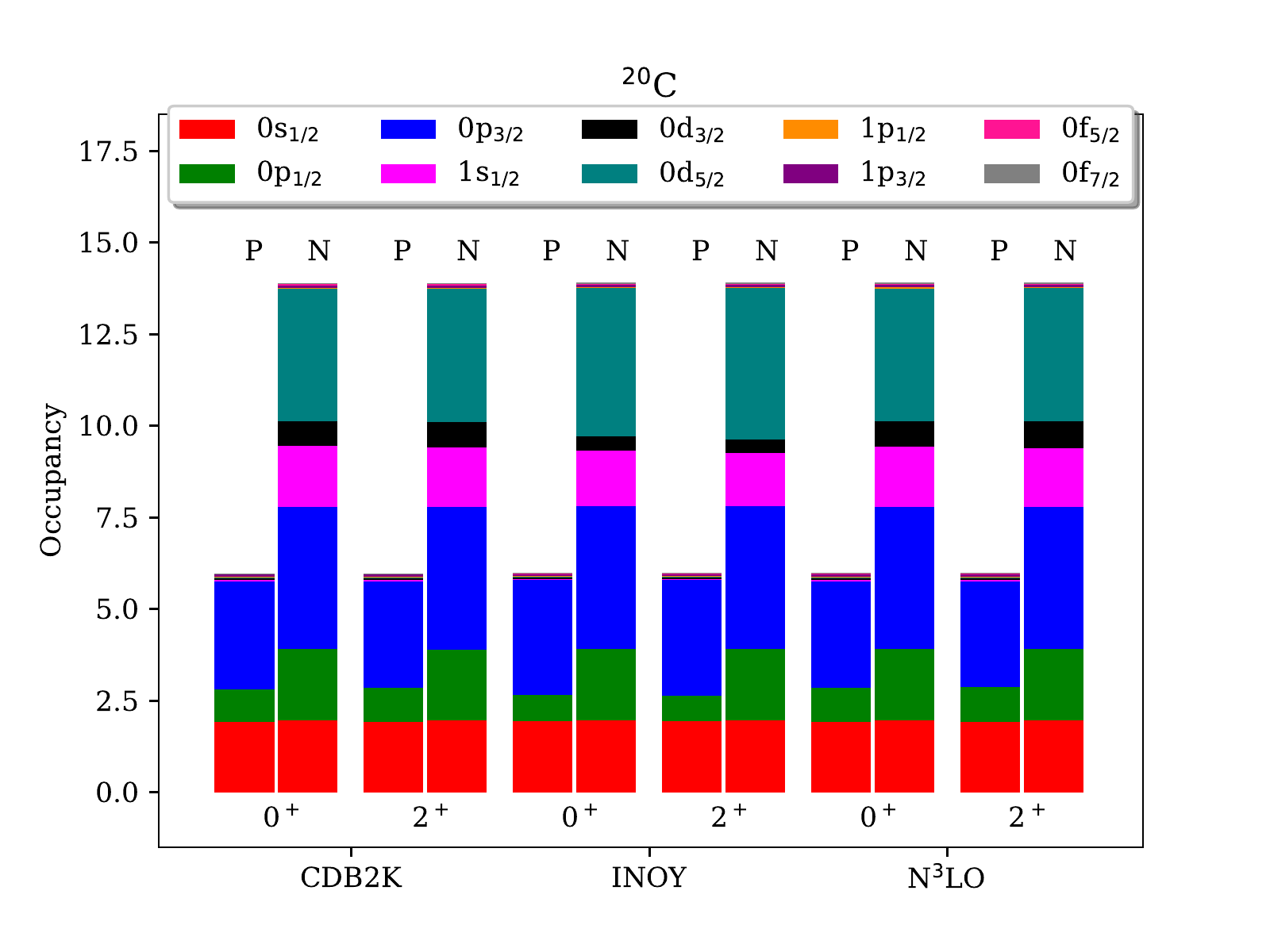}
	\caption{Occupation number for ground and first excited states of $^{18-20}$C isotopes with the CBD2K, INOY  and N$^{3}$LO  interactions. `P' and `N' represent proton and neutron occupation numbers, respectively.}
	\label{Occupancy}
\end{figure*}
Only B(M1) is experimentally known for $^{19}$C corresponding to the transition from $3/2_1^+$ to $1/2_1^+$. For this transition, the experimental value is 0.00321(25) $\mu_N^2$, which is strongly hindered among light nuclei \cite{19C_2015}. In this nucleus, the lowering of the neutron $s_{1/2}$ orbital is responsible for the halo formation, and considerations based on the shell model suggest that the resultant $s_{1/2}$-$d_{5/2}$ degeneracy suppresses the M1 transition. Theoretically obtained values from NCSM calculations also favor this weak M1 transition strength. In calculations with YSOX interaction, we have taken bare g values in Set I, while we have taken effective g value from Ref. \cite{19C_2015} in Set II. In both these cases, B(M1) transition strength is less than 0.001 $\mu_N^2$. Similar calculations have been performed with bare and effective g values for $^{18,20}$C. Experimental data for the quadrupole and dipole moments are not yet available. We have predicted these values using different realistic interactions and compared them with the empirical results. For $^{18}$C,  YSOX interaction gives opposite sign of quadrupole moment over other interactions. Experimental first excited state of $^{19}$C is predicted to be prolate from \textit{ab initio} as well as empirical calculations. Also, the magnitude of quadrupole moments calculated from realistic interactions are comparable to each other. The magnetic moments of g.s. and first excited state is calculated for $^{19}$C and reported in Table \ref{EM_properties}. In the case of $^{20}$C, sign of the magnetic moment is also sensitive to the nuclear interaction. In Ref. \cite{Forssen}, magnetic moment of $^{20}$C with SRG-evolved chiral \textit{NN} interaction is found to be 0.0001(8) $\mu_N$, which is quite small compared to our results.

Further, we have investigated the contribution of different orbitals in the NCSM wave-function for a detailed study of electromagnetic properties in carbon isotopes.
In Fig. \ref{Occupancy}, we have plotted the occupancy of 0$^+$ and 2$^+$ states of even-carbon isotopes $^{18}$C, $^{20}$C and $1/2^+$ and $3/2^+$ of $^{19}$C. The occupation number corresponding to different realistic interactions are compared. Proton and neutron occupation numbers are shown up to the $fp$-shell and beyond which, it is very small, and thus are not shown in the figures. For $^{18}$C, occupancy of $\nu$0d$_{5/2}$ is largest and $\nu$1s$_{1/2}$ is smallest in the case of the  INOY as compared to other interactions for $0^+$ and $2^+$. When we move from 1/2$^+$ to 3/2$^+$ state in $^{19}$C, occupancy of $\nu$1s$_{1/2}$ state significantly increases, while occupancy of $\nu$d$_{3/2}$ and $\nu$d$_{5/2}$ decreases. It is reported in Ref. \cite{20C_2011} that occupation of the 1$s_{1/2}$ orbital leads to the halo structures, hence, our results support the halo structure of $^{19}$C. The $2^+$ state in even carbon isotopes is generated by the excitation of neutrons within $sd$-shell, there is no proton excitation due to large gap between $p_{3/2}$ and $p_{1/2}$ orbitals,  hence, the occupation number of protons for $0^+$ and $2^+$ is almost same.
 Occupancy of $\nu$0d$_{3/2}$ orbital increases for 0$^+$  and $2^+$ states with all interactions as we move from $^{18}$C to $^{20}$C.

\begin{table}[!h]
	\caption{\label{radius}NCSM  ground state point-proton radii for $^{18-20}$C isotopes using CDB2K, INOY and N$^3$LO interactions in the largest basis space compared with the experimental results.}
	\centering
	  \begin{threeparttable}
		\begin{tabular}{cccccc}
			\vspace{-2.8mm}\\
			\hline
			\hline 
			\vspace{-2.8mm}\\
			$ r_{p}$ & Exp.\tnote{a} & Exp.\tnote{b}  &CDB2K & INOY & N$^3$LO  \\
			\hline \vspace{-2.8mm}\\
			
			$^{18}$C & 2.39(4)& 2.42(5)&2.29& 2.06 & 2.30  \\
			$^{19}$C & 2.40(3)&2.43(4) &2.30&2.07  & 2.31  \\
			$^{20}$C & NA & NA &2.30& 2.07 & 2.31 \\
			\hline \hline
		\end{tabular}
	    \begin{tablenotes}
		\item[a]{The data are from Ref. \cite{Kanungo}}
		\item[b] {The data are from Ref. \cite{Tran2}}
	\end{tablenotes}
  \end{threeparttable}
\end{table}

\begin{figure*}
	\includegraphics[width=6cm]{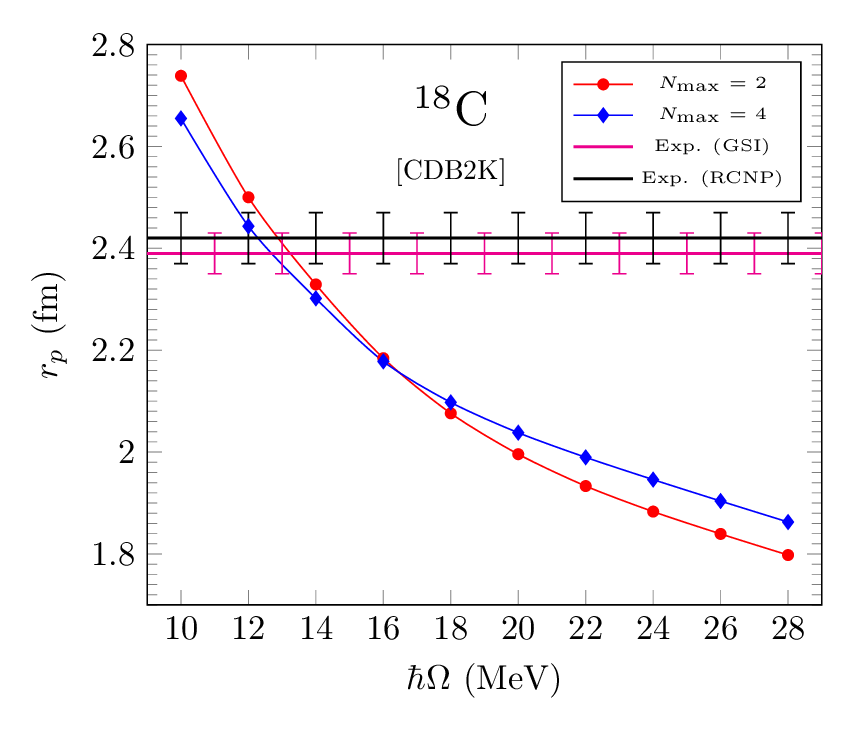}
	\includegraphics[width=6cm]{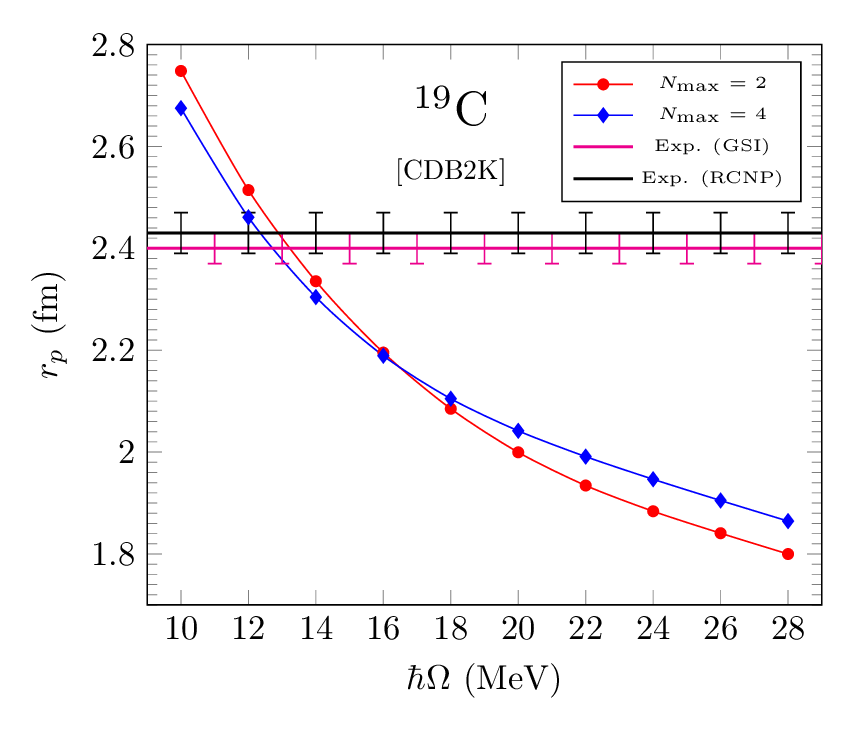}
	\includegraphics[width=6cm]{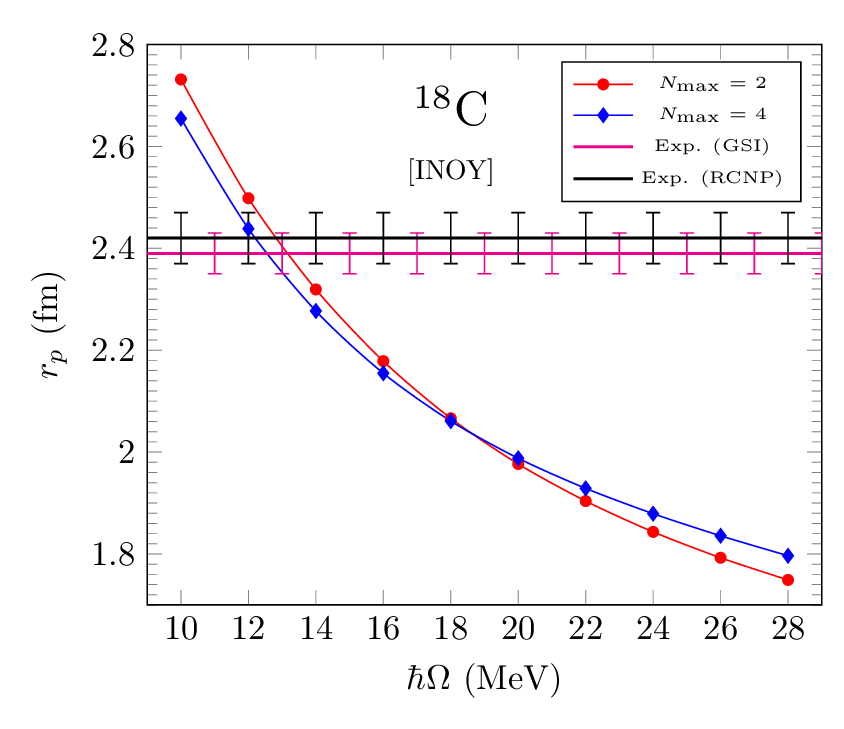}
	\includegraphics[width=6cm]{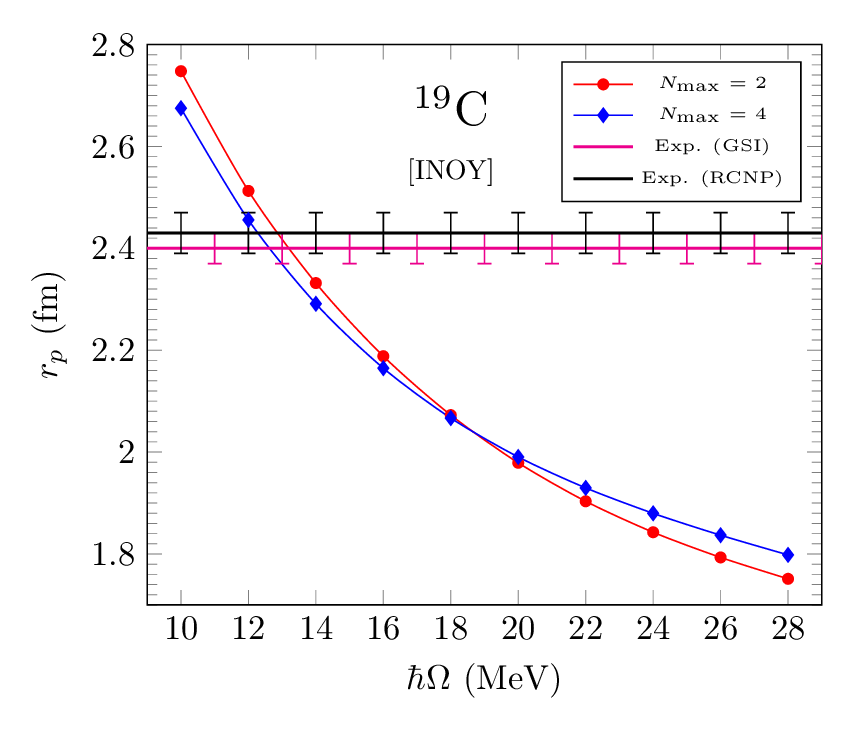}
	\includegraphics[width=6cm]{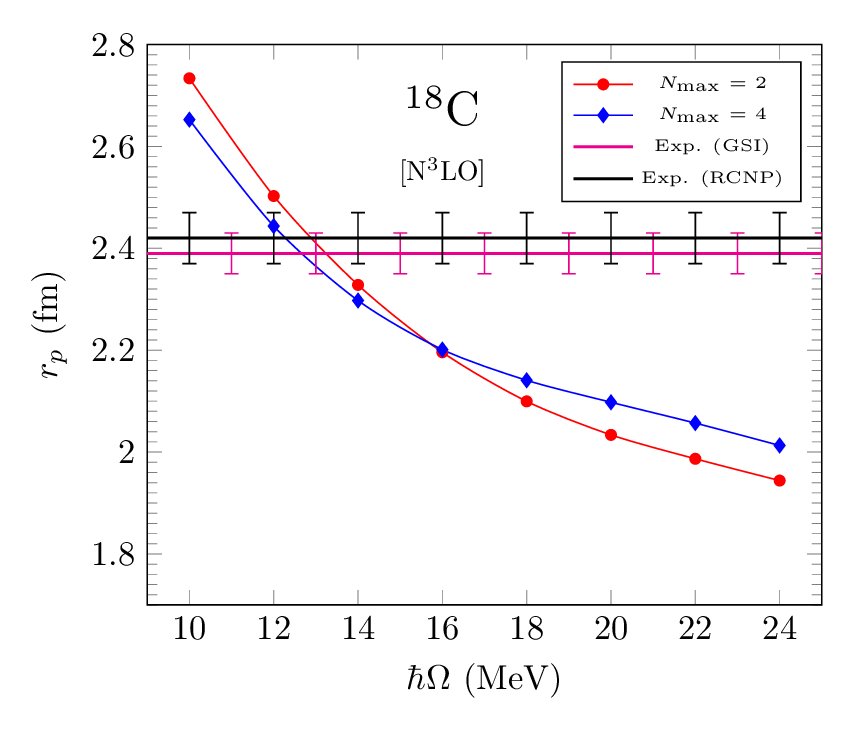}
	\includegraphics[width=6cm]{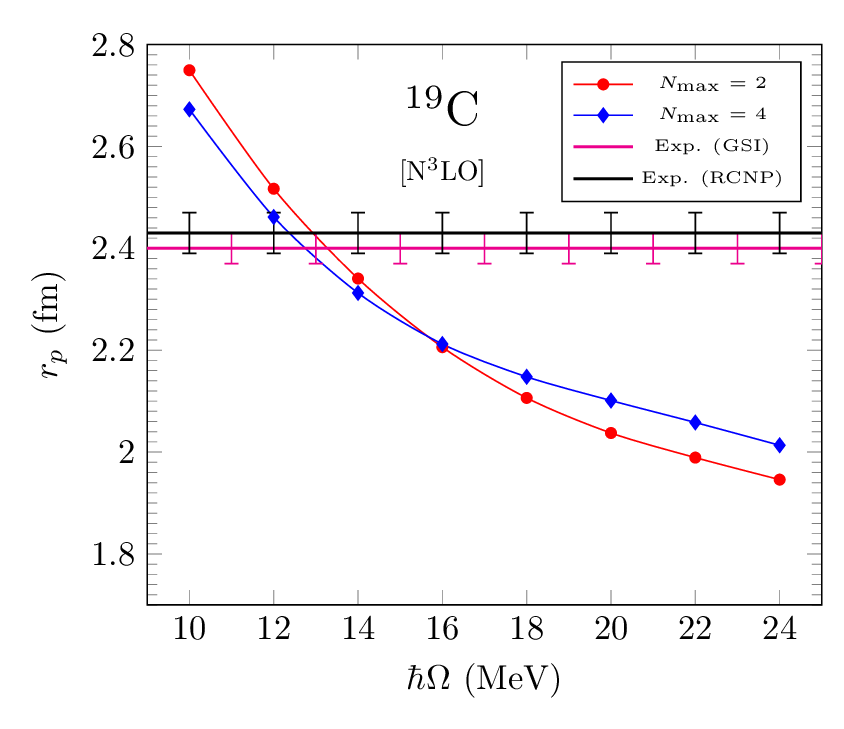}
	\caption{Ground state point-proton radii for $^{18}$C and $^{19}$C with CDB2K, INOY and N$^3$LO interactions as a function of HO frequency corresponding to different $N_{\rm max}$.}
	\label{rp_18,19C}
\end{figure*}

\begin{figure}	
	\includegraphics[width=6cm]{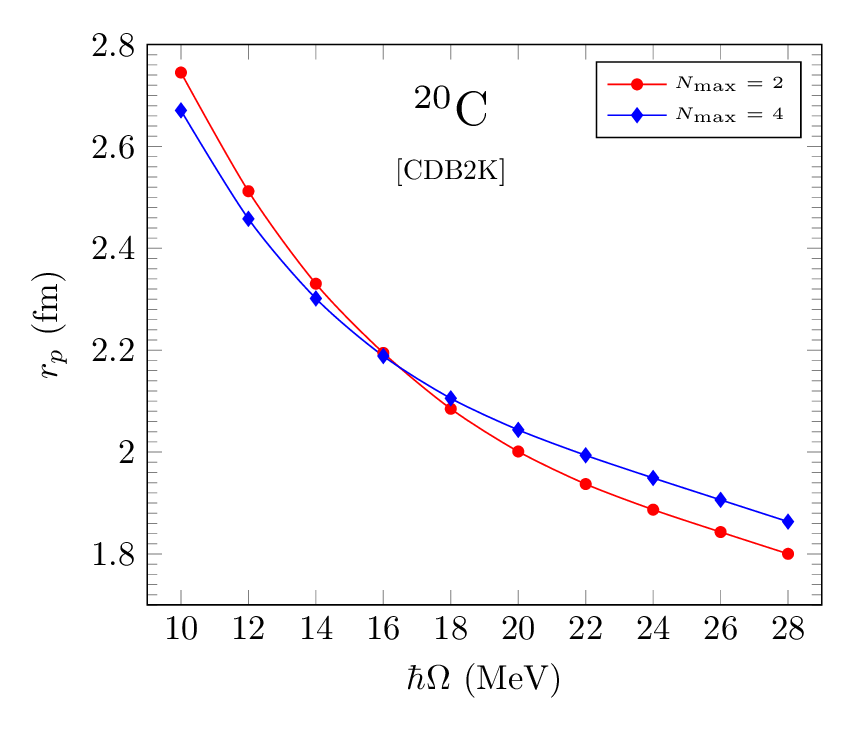}
	\includegraphics[width=6cm]{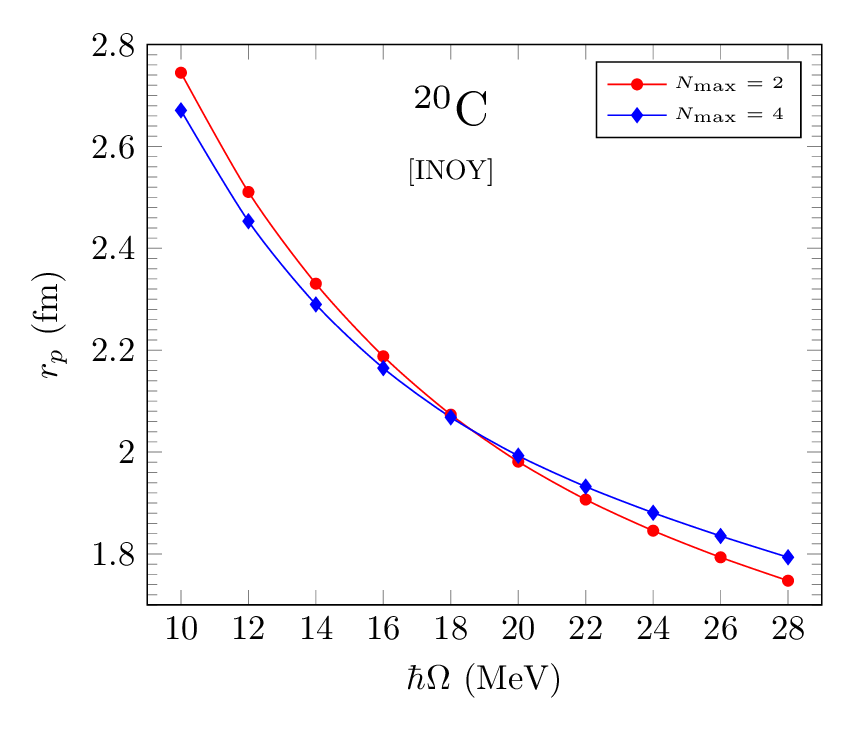}
	\begin{center}
		\includegraphics[width=6cm]{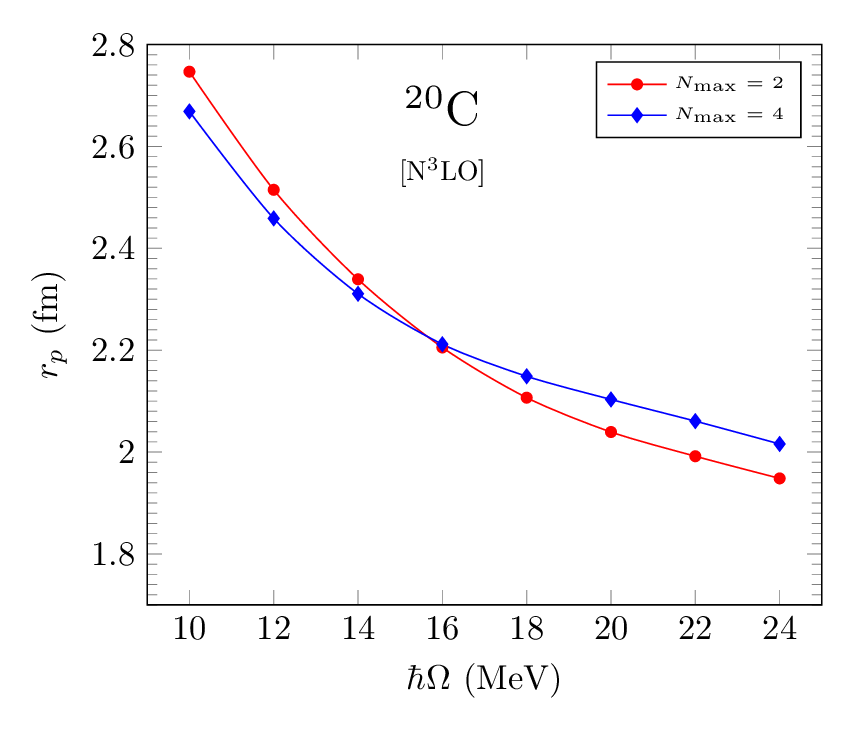}
	\end{center}
	\caption{Ground state point-proton radii for $^{20}$C with CDB2K, INOY and N$^3$LO interactions as a function of HO frequency  corresponding to different $N_{\rm max}$.}
	\label{rp_20C}
\end{figure}

The point-proton radii is an important observable which provides information about the size of the nucleus. In Table \ref{radius}, calculated values of  g.s. $r_p$ using different \textit{ab initio} interactions are reported. These $r_p$ correspond to optimal frequency and largest $N_{\rm max}$. For comparison, experimental data from two different measurements are taken. In Ref. \cite{Tran2}, it is mentioned that $r_p$ are constant throughout the isotopic chain from $^{12}$C to $^{19}$C. Radii obtained from RCNP experiment \cite{Tran2} are $\sim$ 1.3\% larger than radii measured at GSI \cite{Kanungo}. 
When we use N$^3$LO instead of the INOY interaction in NCSM calculations, the radii are found to increase by 12\%.
We can see from Table \ref{radius} that radii obtained from CDB2K and N$^3$LO interactions are very similar, and close to the experimental radii, while INOY underpredicts the $r_p$. Calculated radii for $^{19}$C from realistic interactions are exactly the same as those obtained for $^{20}$C. The almost constant $r_p$ obtained for these isotopes  favors the experimentally observed trend. 

The $r_p$, which is a long-range observable, is sensitive to the larger-$r$ part of the nuclear wave-function, hence, it is challenging to obtain convergent results using NCSM. 
We have examined the dependence of $r_p$ of g.s. for $^{18-20}$C isotopes on HO frequency and $N_{\rm max}$ as shown in Fig. \ref{rp_18,19C}-\ref{rp_20C}. We have also shown the experimental data obtained from \cite{Kanungo,Tran2} measurements by the horizontal line with uncertainties. In Refs. \cite{Caprio,Shirokov}, it is reported that radii curves corresponding to different $N_{\rm max}$ as a function of $\hbar$$\Omega$ intersect at a common point and this point gives the converged radii.  In our case, there are only two: $N_{\rm max}$ = 2 and $N_{\rm max}$ = 4. Thus, we have estimated the crossing point of these two curves as the converged radii. For $^{18}$C, the obtained converged radii are 2.16, 2.03 and 2.21 fm corresponding to CDB2K, INOY and N$^3$LO interactions, respectively. We note that the calculated $r_p$ decreases with increasing $N_{\rm max}$ at frequencies lower than the crossover point and it increases with $N_{\rm max}$ at frequencies higher than the crossover frequency. In a similar way, from the figures we can obtain converged $r_p$ for $^{19}$C and $^{20}$C corresponding to different interactions. We can say that to obtain true converged radii, one needs an increment in the basis size beyond $N_{\rm max}$ = 4.

\section{Summary and Conclusions}
In the present work, we have studied nuclear structure properties of neutron rich carbon isotopes $^{18-20}$C within the \textit{ab initio} NCSM method using CDB2K, INOY  and N$^3$LO realistic \textit{NN} interactions. We have calculated low-lying energy states, electromagnetic properties and point-proton radii for $^{18-20}$C isotopes. Conventional shell model calculations with the YSOX interaction are also performed for comparison. We found that binding energy obtained with INOY interaction is closest to the experimental data among three realistic interactions, still, it is far from the experimental data. For further improvement, calculations with an enlarged basis size are needed. 
It is also noted that the chiral interaction used is not state-of-the-art any more. To get more accurate results, we need to include N$^4$LO  interaction and beyond in our calculations. By comparing the NCSM B(E2) results with the experimental data, we find that the obtained B(E2) values of $^{18}$C and $^{20}$C are highly sensitive to the nuclear interaction. We have predicted ground and first excited states quadrupole and magnetic moments for these isotopes where experimental data are not yet available. Thus our present systematic study will add more information to earlier work \cite{Forssen}. More experimental studies for energy spectra, B(E2), electric quadrupole and magnetic dipole moments for these nuclei are required for a complete nuclear structure description. Our theoretical prediction of these nuclei will be helpful in the future for comparison with the experimental data.

\section*{Acknowledgement}

We acknowledge a research grant from SERB (India), CRG/2019/000556. 
P.C. acknowledges financial assistance from MHRD (Government of India) for her Ph.D. thesis work.
 We would like to thank Prof. Petr
Navr\'{a}til for providing us his \textit{NN} effective interaction code. 
We also thank Prof. Christian Forss\'en for making available the pAntoine code. We acknowledge National Supercomputing Mission (NSM) for providing computing resources of `PARAM Ganga' at Indian Institute of Technology Roorkee, which is implemented by C-DAC and supported by the Ministry of Electronics and Information Technology (MeitY) and Department of Science and Technology (DST), Government of India.

\section*{References:}

\bibliographystyle{utphys}
\bibliography{references}

\end{document}